\documentclass[aps,prl,twocolumn,showpacs,groupedaddress]{revtex4}
\usepackage[pdftex]{graphicx}
\usepackage{psfrag}
\usepackage{color}
\usepackage{amsbsy}
\usepackage{amssymb}

\def\a{\alpha}
\def\b{\beta}

\def\dd{\mbox{d}}

\begin{document}
\bibliographystyle{apsrev}


\title{Peeling and Sliding in Nucleosome Repositioning}


\author{Tom Chou}
\affiliation{Dept. of Biomathematics \& Dept. of Mathematics, UCLA, Los Angeles, CA 90095}


\date{\today}

\begin{abstract}
We investigate the mechanisms of histone sliding and detachment with a
stochastic model that couples thermally-induced, passive histone
sliding with active motor-driven histone unwrapping.  Analysis of a
passive loop or twist defect-mediated histone sliding mechanism shows
that diffusional sliding is enhanced as larger portions of the DNA
is peeled off the histone.  The mean times to histone detachment and
the mean distance traveled by the motor complex prior to histone
detachment are computed as functions of the intrinsic speed of the
motor.  Fast motors preferentially induce detachment over
sliding. However, for a fixed motor speed, increasing the histone-DNA
affinity (and thereby decreasing the passive sliding rate) {\it
increases} the mean distance traveled by the motor. 
\end{abstract}



\maketitle


The nucleosome is comprised of double-stranded DNA that wraps
1${3\over 4}$ turns around the edge of disk-shaped histone
proteins. These structures impart a DNA accessibility-based code in
addition to the sequence-based genetic code \cite{WIDOM}.  Positioning
of the histone particles depends on the local histone-DNA affinity
(via {\it e. g.}, DNA bendability \cite{WIDOM}) and can modulate the
accessibility of DNA.  The remodeling of nucleosomes, involving
perhaps histone sliding and/or detachment occurs during cellular
processes that require DNA accessibility \cite{CAIRNS}.  For example,
in replication, translation of a replication fork along the DNA
requires histones be either shifted or removed \cite{HS,IMB}. It has
been proposed that these processes may be passive, in which parts of
the histone-DNA complex thermally unwind and rewind, leaving transient
regions of the DNA accessible
\cite{WIDOM1,CAIRNS,HS,PB,GERLAND}. Histone sliding has also been
observed and modeled theoretically \cite{SLIDE0}.  However, since the
histone-dsDNA binding energy is $\sim 40k_{B}T$, sliding mechanisms
must exploit low energy thermal excitation such as the propagation of
extra-length loops or twist defects \cite{SLIDE1}. Although these
thermal mechanisms result in slow histone sliding, more rapid
nucleosome remodeling can be catalyzed by molecular motor enzymes
\cite{OWEN,TYLER}.  A number of ATP-dependent remodeling factors, such
as SWI/SNF, ISWI, and SWI2/SNF2 have been identified to be important
during transcription \cite{CLS,PB,IMB}. 
The action of these remodeling factors may be required for more rapid
detachment of histones from the DNA substrate. Apart from the
observation that chromatin remodeling factors have conserved
helicase-like ATPase domains \cite{CLS}, and that histones can both
slide along and dissociate from DNA, there has been no explicit
mechanistic hypothesis for enzymatically mediated chromatin
remodeling.


In this Letter, we propose and explore the consequences of a simple,
 explicit mechanism for enzyme-mediated chromatin remodeling.  The
 remodeling factor is modeled as a processing motor that runs into an
 isolated dsDNA-wrapped histone particle. Such a motor not only can
 bias histone sliding, inducing a drift, it can also wedge itself
 underneath the histone, peeling off the DNA.  The full stochastic
 process is described by an effective master equation, from which we
 find the mean times to histone detachment, and the mean travel
 distance of the motor before it peels the histone off the DNA.
\begin{figure}[t]
\begin{center}
\includegraphics[height=8.0cm]{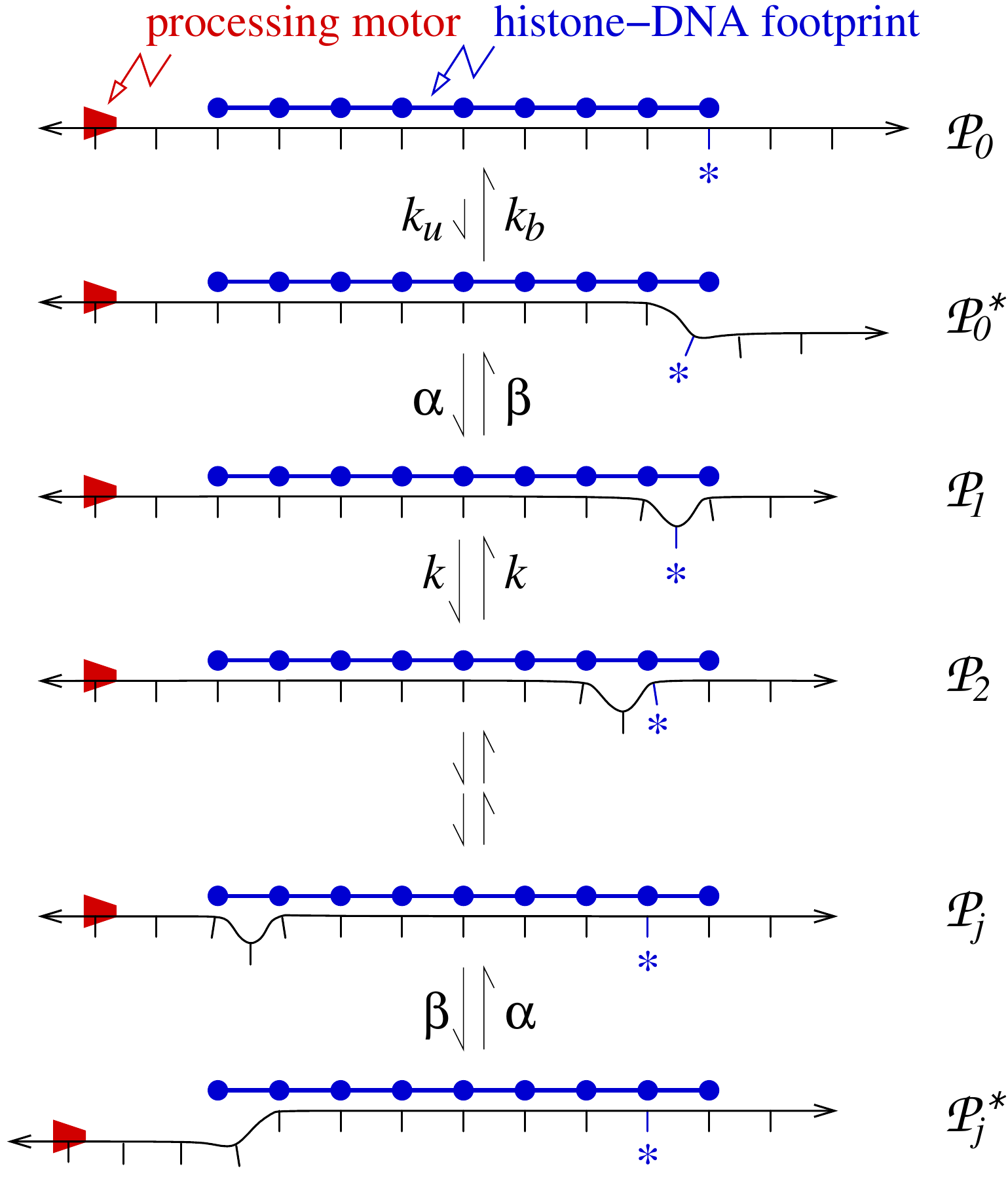}
\end{center}
\vspace{-4mm}
\caption{The kinetics of single, isolated histone motion mediated by
  thermal creation and propagation of loops. Flaps unbind and bind at
  either end of the contact region with rate $k_{u}$ and $k_{b}$,
  respectively. Closure of a flap into a loop occurs at rate $\alpha$
  and loop ejection into a flap occurs with rate $\beta$. Unbiased
  loop hopping rates in the interior of the contact region occurs with
  rate $k$. The asterisk labels a fixed site on the DNA substrate.  A
  motor complex is shown to the left.}
\label{FIG1}
\end{figure}
Figure \ref{FIG1} depicts the histone-DNA contact region, or
footprint, and a thermal inchworming mechanism.  Small flaps
(or defects) are thermally excited with rate $k_{u}$ by fluctuating
segments of DNA that momentarily unwrap from the either edge of the
histone-DNA contact region. The flap can rebind to the segment from
which it detached with rate $k_{b}$, or, it can reattach to the
next-nearest neighbor segment with rate $\a$, generating a loop.  Once
an interior loop is formed, it can hop to the right or left with rate
$k$. The histone has translated one loop arclength only when a loop
entering on one side of the contact region has left the opposite end.
In this reptation-like mechanism, the net motion of an isolated
histone is related to the statistics of diffusing loop ``particles''
that exit the end opposite from which they were thermally excited,
analogous to the gambler's ruin problem \cite{REDNER}. The sites of
the coarse-grained lattice represent individual histone-DNA contact
``bond'' segments, separated by a distance equivalent to the typical
arclength of a flap, about 10 base pairs \cite{HS}.

For a histone particle wrapped with $n$ segments containing a single
interior loop, we denote the probability that the loop is at position
$i$ ($0\leq i \leq n$) by ${\cal P}_{i}$.  Since the injection of
loops is slow ($\a \ll k_{b}$) \cite{SLIDE1,JMB}, the probability for
multiple loops is low, and loop-loop interactions can be neglected.
The master equation governing the interior diffusive motion of the
loop is $\dot{{\cal P}}_{i} = k {\cal P}_{i-1}+k{\cal P}_{i+1}-2k{\cal
P}_{i}$ for $2 \leq i \leq n-1$. Near the left edge the boundary
conditions are $\dot{{\cal P}}_{1} = k{\cal P}_{2}-(\b+k){\cal
P}_{1}+\a {\cal P}_{0}^{*}$ and $\dot{{\cal P}}_{0}^{*} = k_{u}{\cal
P}_{0}-(k_{b}+\a){\cal P}_{0}^{*} + \b {\cal P}_{1}$. Similar
expressions hold at the right edge.  The probability that a loop exits
the opposite end from which it entered can be found by considering a
unit source of probability ${\cal P}_{0}^{*} = 1$, and solving the
resulting recursion equations \cite{REDNER}.  The total probability of
starting in state ${\cal P}_{0}^{*}$ and arriving in state ${\cal
P}_{n}$ is ${\a k \over (n-1)\b^{2} k_{b} + \beta k(\a+2k_{b})}$.
Since flaps are thermally generated with rate $k_{u}$, and
produce the state ${\cal P}_{n}^{*}$ with rate $\beta {\cal P}_{n}$,
the effective hopping rate of the entire histone along the DNA is

\begin{equation}
p_{n} = {\a k k_{u} \over (n-1)\b k_{b} + k(\a+2k_{b})}.
\end{equation}

\noindent For an isolated, fully-wrapped histone, flap generation at
each end is statistically independent and its motion is that of
unbiased diffusion with diffusivity $p_{N}$. The loop hopping rate $k$
will be approximated as two sequential flap detachment and
reattachment steps, without any additional barriers, giving $k \sim
k_{u}k_{b}/(k_{u}+k_{b})$. Similarly, $\b$ can be approximated by the
detachment rate. Finally, the insertion of a loop is proportional to
the flap reattachment rate, but reduced by a factor associated with
the bending required to form a loop. Thus, we use the physical rate
estimates

\begin{equation}
\a \approx \a_{0}k_{b}, \quad \b \approx k_{u}, \,\,\, \mbox{and} \,\,\,
k \approx k_{u}k_{b}/(k_{u}+k_{b}), 
\label{ESTIMATES}
\end{equation}

\noindent where $\a_{0} \sim e^{-E_{b}/k_{B}T} \ll 1$ scales the
loop injection rate by the energy $E_{b}$ required to bend the DNA
into a small loop.  These approximations make our results depend on
just $w/\alpha$ and $K=k_{b}/k_{u}$, and the maximum number $N$ of
DNA-histone contacts. Since $E_{b} \sim 23k_{B}T$ \cite{SLIDE1,JMB}
and $\a_{0} \sim 10^{-10}$, we present all results in terms of
quantities rescaled to the slowest rate $\alpha$ in the problem.

\begin{figure}[h]
\begin{center}
\includegraphics[height=4.2cm]{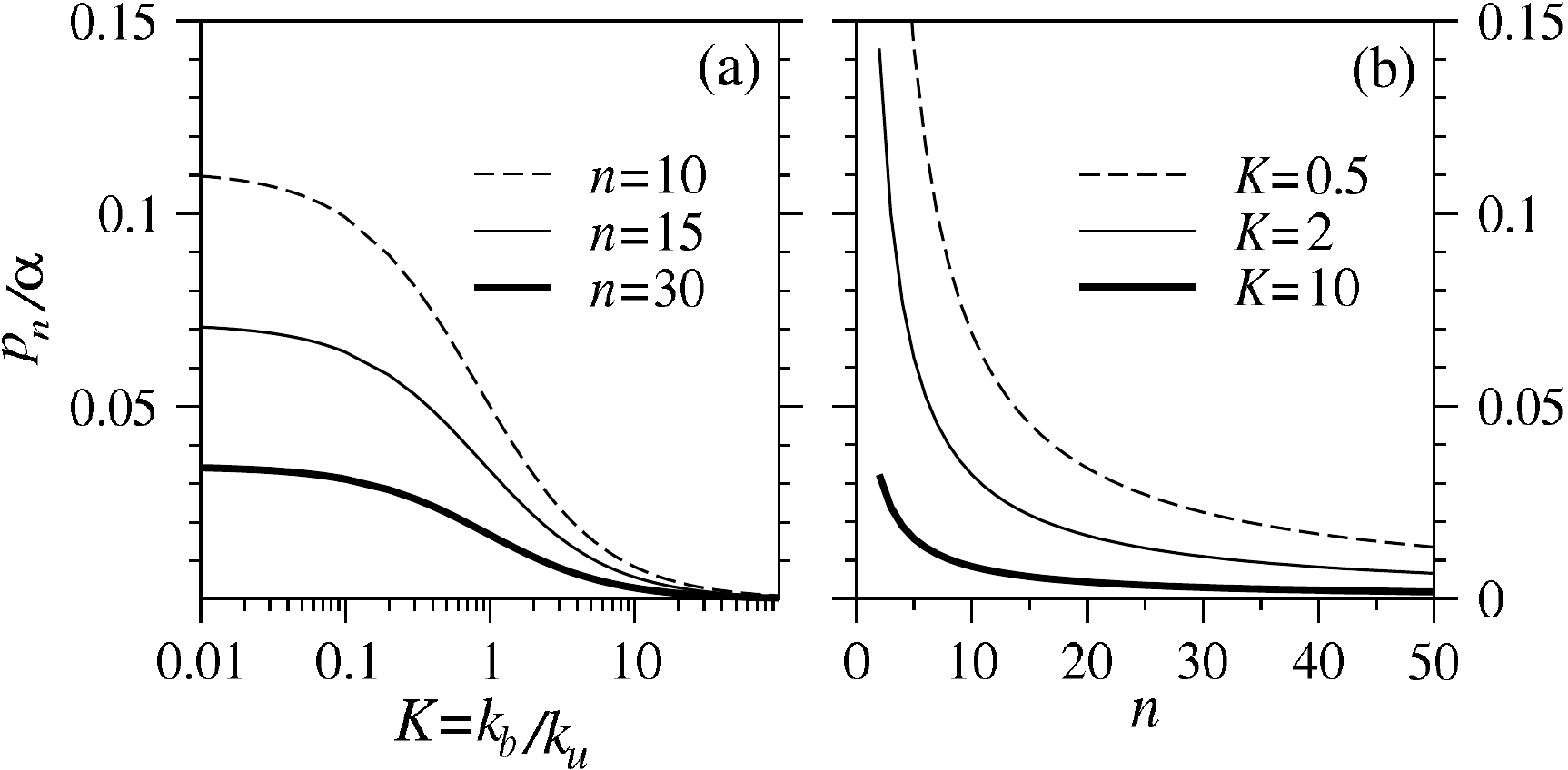}
\end{center}
\vspace{-4mm}
\caption{(a) Effective diffusion coefficient $p_{n}/\alpha$ as a
function of flap binding constant $K=k_{b}/k_{u}$. (b) Diffusivity as
a function of the degree of histone-DNA wrapping $n$. The parameters
used are $\a = 10^{-10}k_{b}, \b = k_{u}$, and
$k=k_{b}k_{u}/(k_{b}+k_{u})$ (Eq.  \ref{ESTIMATES}).}
\label{PN}
\end{figure}

Figure \ref{PN}a shows the normalized effective hopping rate
$p_{n}/\alpha$ as a function of $K=k_{b}/k_{u}$ and $n$.
For fixed $k_{b}$, $p_{n}$ decreases with decreasing flap unbinding
rate $k_{u}$. 
The effective hopping rate also depends on the number $n$ of
histone-DNA contact elements, as shown in Fig. \ref{PN}b.  The higher
the degree of wrapping of a histone, the larger the $n$, and the
smaller its loop-mediated diffusion coefficient.



Now consider a processing motor complex that moves unidirectionally
with rate $w$ along the DNA. For simplicity, we assume that the motor
moves by steps of length approximately equal to that of a thermally
generated loop.  Fig. \ref{FIG1} shows a motor to the left of a
histone-DNA contact region.  The motor can advance only if the segment
in front of it is cleared and not attached to the histone.  The edge
of the contact region just ahead of a motor can be cleared by a
thermally-excited flap, allowing the motor to possibly move forward
and slip under it. Consequently, the effective forward hopping rate
$\bar{w}$ of a motor that is peeling a histone is reduced from the
free hopping rate $w$ by a factor dependent on the reattachment rate
$k_{b}$:

\begin{equation}
\bar{w} = {w k_{u} \over w + k_{b}}.
\end{equation}

The motor can also advance into the contact region by loops generated
at the {\it opposite} end, but that have arrived after diffusing
across the footprint.  Thermal excitation of a flap at the far end,
and subsequent propagation of a loop to the opposite end, occurring
with rate $p_{n}$, momentarily creates a flap of two bond segments
near the motor.  Again, flap closing competes with motor insertion
underneath the flap.
\begin{figure}
\begin{center}
\includegraphics[height=4.2cm]{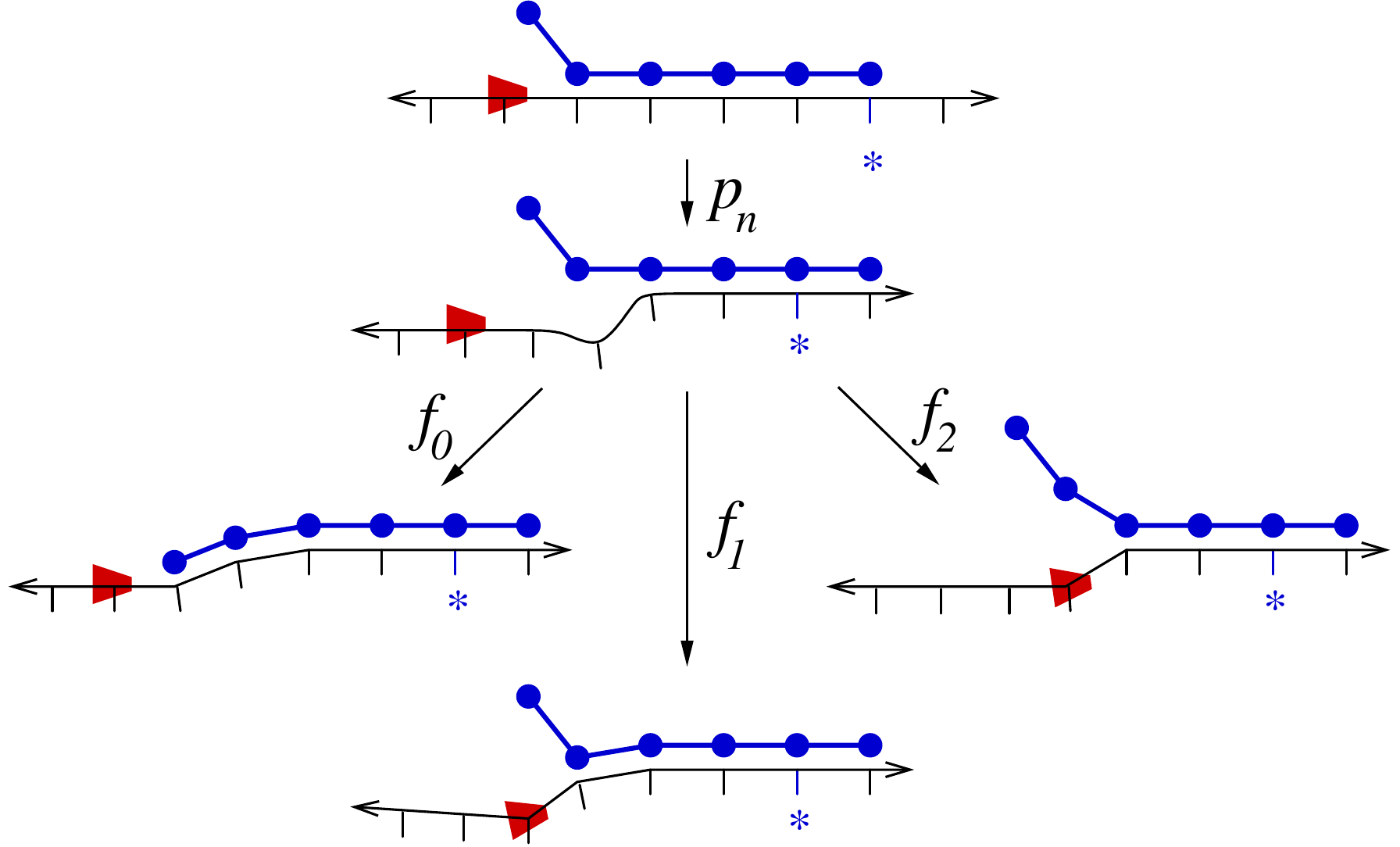}
\end{center}
\vspace{-4mm}
\caption{Details of the motor-flap interaction. When a loop generated
on the opposite end exits the side nearest the motor (with effective
rate $p_{n}$), the motor can move forward 0,1, or 2 steps
with relative probabilities $f_{0}, f_{1}$, and $f_{2}$, respectively.}
\label{PATHS}
\end{figure}
As shown in Fig. \ref{PATHS}, there are three fates available to a
two-bond flap nearest the motor. The motor can remain in its position
and both bonds of the flap can close (with effective rate $k_{b}/2$),
the motor can move one position forward and one bond of the flap can
close (with effective rate $wk_{b}/(w+k_{b})$), or, the motor can move
forward two sites (with effective rate $w/2$) under the flap,
preventing it from closing at all. The total transition rate available
to the two-bond flap is $r_{T}=k_{b}/2 + k_{b}w/(k_{b}+w) + w/2$, from
which we find the probabilities for each pathway

\begin{equation}
f_{0} = {k_{b}\over 2r_{T}},\,\, f_{1} = {k_{b}w \over r_{T}(k_{b}+w)},
\,\, \mbox{and} \,\, f_{2} = {w \over 2r_{T}}.
\end{equation}

\noindent Note that $f_{0}+f_{1}+f_{2} = 1$ and that the ratios of
these pathways are determined by the motor speed $w$.  These
probabilities apply only to flaps on the motor side of the contact
region. On the opposite, motor-free end, we assume that single bond
flaps close sufficiently fast ($K\equiv k_{b}/k_{u} \gg 1$) such that
spontaneous desorption of histones is negligible in the absence of a
processing motor.


As the progressing motor partially unwraps the histone, the length $n$
of the contact region decreases, increasing the effective sliding rate
$p_{n}$ of the remaining contact region (cf. Fig. \ref{PN}b).  The
sliding mechanism becomes increasingly important as the degree of
unpeeling increases, and the histone becomes more efficient at
diffusively escaping the motor. If $m$ is the position of the motor,
and $n$ is its distance to the far end of the footprint, a master
equation that incorporates all of the kinetics described above can be
written for the probability $P(m,n,t)$:

\begin{widetext}
\begin{equation}\begin{footnotesize}
\begin{array}{ll}
\displaystyle \dot{P}(m,n,t) & = p_{N}\left[P(m,n+1)
-2P(m,n) + P(m,n-1)\right]+w\left[P(m-1,n+1)-P(m,n)\right],\quad n\geq N+1 \\[13pt]
\displaystyle \dot{P}(m,N,t) & = -(\bar{w}+p_{N})P(m,N) +wP(m-1,N+1) + f_{0}p_{N-1}P(m,N-1)
+ f_{1}p_{N}P(m-1,N) +p_{N}P(m,N+1),\\[13pt] 
\displaystyle \dot{P}(m,n,t) & 
= -(\bar{w}+p_{n})P(m,n) + \bar{w}P(m-1,n+1) +
f_{0}p_{n-1}P(m,n-1) + f_{1}p_{n}P(m-1,n) \\[13pt]
\: & \hspace{7.4cm} + f_{2}p_{n+1}P(m-2,n+1), \,\quad 3\leq n < N,
\label{FULLEQNS}
\end{array}\end{footnotesize}
\end{equation}
\end{widetext}

\noindent with $\dot{P}(m,1) = -k_{u}P(m,1) +
\bar{w}P(m-1,2)+f_{2}p_{2}P(m-2,2)$ and $\dot{P}(m,2) =
-(\bar{w}+p_{2})P(m,2)+\bar{w}P(m-1,3)+f_{1}p_{2}P(m-1,2)+
f_{2}p_{3}P(m-2,3)$. The statistics of the histone detachment time can
be found by considering only the relative distance $n$, and when it
first reaches $n=0$.  Upon summing the time-integrated probabilities
over all possible motor positions $m\geq 0$, Eqs. \ref{FULLEQNS} are
succinctly expressed as

\begin{equation}
a_{n+1}Q_{n+1}-(a_{n}+b_{n})Q_{n}+b_{n-1}Q_{n-1} = -\delta_{n,N}
\label{QEQNS}
\end{equation}

\noindent where $Q_{n} \equiv \int_{0}^{\infty}\dd t
\sum_{m=0}^{\infty} P(m,n,t)$ and $\delta_{n,N} = \sum_{m=0}^{\infty} P(m,n,0)$ is the
initial condition with the motor  at the left end of the contact
region of a fully wrapped histone. In Eq. \ref{QEQNS}, the transition 
rates are

\begin{equation}
\begin{array}{lll}
a_{n}=w+p_{N}, &  b_{n}=p_{N}  &\quad  n>N \\[13pt]
a_{n} = \bar{w}+f_{2}p_{n}, &  b_{n} = f_{0}p_{n} & \quad 2\leq n\leq N.
\end{array}
\end{equation}

\noindent Conditions just prior to detachment $(n=1)$ also require
$b_{0}=b_{1}=0$ and $a_{1}=k_{u}$.

The mean first detachment time of the process
described by Eq. \ref{QEQNS} is found by solving the corresponding
Backward Kolmogorov equation \cite{REDNER}, or by directly
time-integrating the solution $\langle T_{d}\rangle =
\sum_{n=1}^{\infty}\int_{0}^{\infty}Q_{n}(t)\dd t$:

\begin{equation}
\langle T_{d}\rangle = \sum_{n=1}^{N} {1 \over a_{n}} +
\sum_{n=1}^{N}\sum_{i=1}^{\infty} {1\over a_{n+i}}\prod_{k=0}^{i-1}
{b_{n+k} \over a_{n+k}}.
\label{MFPT0}
\end{equation}


The mean distance $\langle m_{d}\rangle$ traveled by the 
motor before it  detaches the histone  can be
constructed from the time-integrated quantity $R_{n} \equiv
\int_{0}^{\infty}\dd t \sum_{m=0}^{\infty}mP(m,n,t)$, which obeys an
equation similar to (\ref{QEQNS}):
$a_{n+1}R_{n+1}-(a_{n}+b_{n})R_{n}+b_{n-1}R_{n-1} = G_{n}$.
The additional inhomogeneity terms $G_{n>N}=-Q_{n}$, $G_{N} =
-wQ_{N+1}$, and $G_{n\leq N} =
-(\bar{w}+2f_{2}p_{n+1})Q_{n+1}-f_{1}p_{n}Q_{n}$, rely on the solution
$Q_{n}$ to Eq. \,\ref{QEQNS}. Upon solving for $R_{n}$, we find

\begin{equation}
\begin{array}{l}
\langle m_{d}-1\rangle = \displaystyle k_{u}\int_{0}^{\infty}\!\!\dd t
\sum_{m=0}^{\infty}mP(m,1,t) = k_{u}R_{1}.
\end{array}
\end{equation}

\begin{figure}[ht]
\begin{center}
\leavevmode
\includegraphics[height=4.2cm]{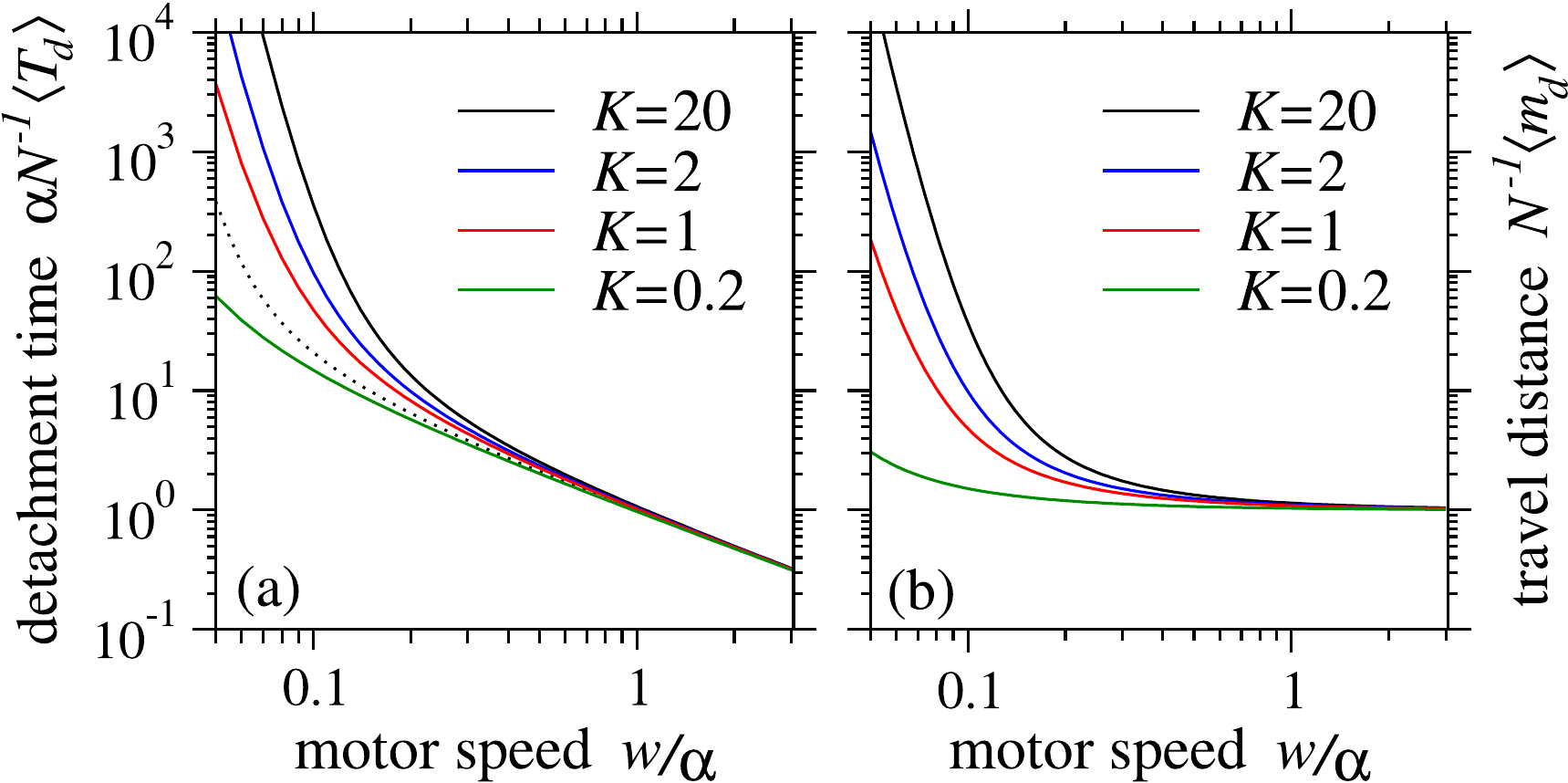}
\end{center}
\vspace{-4mm}
\caption{(a) The normalized mean time to detachment, $\alpha\langle
T_{d}\rangle/N$, as a function of the normalized motor speed
$w/\alpha$, for various $K$.  
The dotted
curve shows the dramatic decrease in detachment times corresponding to
$K=20$ and a hypothetically fixed small diffusivity $p_{N}$ that
suppresses sliding. (b) The mean distance $\langle m_{d}\rangle$
traveled by the motor before it peels off the histone.  In both plots,
the maximum wrapping length was set to $N=15$, corresponding to a
flap/loop length of about 10 base pairs. Since we have assumed tight
binding (large $K$), and ignored spontaneous histone desorption, our
results $\langle T_{d}\rangle$ and $\langle m_{d}\rangle$ for smaller
values of $K$ are only upper bounds.}
\label{MFPT}
\end{figure}

In Fig. \ref{MFPT}a we plot the mean first detachment times as a
function of the motor speed, for $N=15$ and various flap binding
constants $K=k_{b}/k_{u}$. For slow motors,
the double flaps near the motor nearly always close ($f_{0}
\rightarrow 1$) before the motor can advance.  The histone unit
diffuses along the DNA with rate $p_{N}$ and is only occasionally
blocked by a relatively stationary motor. It spends most of its time
away from the motor, preventing it from slipping under flaps that
momentarily open.  The resulting mean first detachment time
diverges exponentially in the $w/\alpha \rightarrow 0$ limit.

In the large $w/\a$ limit, the remodeling complex slips under the
flaps before loops can propagate through the contact region and the
peeling mechanism dominates.  As $w$ is increased, the peeling time
decreases. In the extremely fast limit $w/k_{b} \gg 1$, the motor
nearly always inserts itself underneath flaps before they can reattach
($f_{0}\approx f_{1}\approx 0$, and $f_{2} \approx 1$) and
Eq. \ref{MFPT0} simplifies to $\langle T_{d}\rangle = k_{u}^{-1} +
\sum_{n=2}^{N}(k_{u}+p_{n})^{-1} < Nk_{u}^{-1}$. The mean detachment
time of a histone subjected to an infinitely fast motor would be
$\langle T_{d}\rangle = N/k_{u}$ if flaps originated only in front of
the motor.  However, flap generation at, and loop propagation from the
other end of the contact region enhances the arrival of flaps near the
motor, thereby decreasing $\langle T_{d}\rangle$ to values below
$N/k_{u}$.





In Fig. \ref{MFPT}b, we plot the mean distance traveled by the motor
before it detaches a histone, as a function of $w/\alpha$ for various $K$. For
low motor speeds, sliding dominates peeling, and the motor moves
significantly.  Note as the flap binding affinity $K$ increases, the
mean distance traveled by the motor also {\it increases}.  For fixed
$k_{b}$, increasing $K$ by decreasing $k_{u}$ leads to a dramatic
increase in $\langle T_{d}\rangle$, especially at low motor speeds.
This extra time allows the histone to slide a longer distance before
being peeled off by the motor.  Although increasing $K$ reduces the
effective sliding rate of an isolated histone, mechanistically, the
increase in mean distance traveled arises from the relative
suppression of double flaps near the motor, slowing down the peeling
process.

For larger $w/\a$, peeling becomes dominant and the motor moves a
mean distance $\langle m_{d}\rangle \approx N$ before scrapping the
entire histone off. Again, in the $w/k_{b} \gg 1$ limit,
an analytic form can be explicitly found: $\langle m_{d} \rangle/N =
1+\sum_{n=2}^{N}p_{n}/(k_{u}+p_{n})$.
This slow increase in the {\it normalized} mean distance traveled,
$\langle m_{d}\rangle/N$, with $N$ arises from the relative reduction
of double flaps near the motor, thereby diminishing the frequency of
this efficient mode of peeling.  This reduction occurs for larger $N$
through the loop propagation probability $p_{n}$.

In summary, we have presented and analyzed a stochastic model for
ATP-dependent chromatin remodeling that incorporates a competition
between sliding and peeling of a single isolated histone. Our main
findings include formulae for the mean detachment time and travel
distance, as functions of motor speed and binding affinity.  Histones
with larger binding affinity to the DNA substrate slide farther before
being peeled off by the motor.  We have neglected effects such as DNA
sequence dependence (which might affect $w, k_{u}, k_{b}$, and $k$)
and twist defect mobility \cite{HS}. This latter mechanism may only
open much smaller end flaps that allow motor insertion and peeling.
Nonetheless, our results are useful for dissecting and quantifying
{\it in vitro} measurements of passive sliding and ATPase-assisted
nucleosome remodeling, particularly when physical chemical conditions
can be controlled and $w, k_{u}, k_{b}$ tuned.

Nucleosomes {\it in vivo} typically consists of an array of
interacting histones \cite{TYLER,TONKS}.  Thus, our results would be
valid only if the typical DNA linker length $\ell$ between two
adjacent histones were larger than $\langle m_{d}\rangle + N$. If
$\ell \leq \langle m_{d}\rangle + N$, the histones would be squeezed
together by the motor. In such cases, peeling would be enhanced as the
sliding mobility is reduced due to histone-histone exclusion.
Finally, it should be mentioned that the analysis of
Eq.\,\ref{FULLEQNS} is directly applicable to the study of
dehybridization and driven translocation of double-stranded,
mononucleotide nucleic acids through nanopores \cite{BB}. Our results,
after reidentification of the parameters, provide an analytic solution
to the mean passage times of translocation in the high driving force
limit.

\vspace{3mm}

This work was supported by the NSF (DMS-0349195) and by the NIH
(K25AI41935).



\end{document}